\input{aipcheck}

\documentclass[
    ,final            
  ]
  {aipproc}

\layoutstyle{6x9}

\begin{document}

\title{Is room-temperature superconductivity with phonons possible?}

\classification{74.70.-b; 71.10.-w; 71.10.Hf; 71.10.Li;}
\keywords      {Bose-Einstein condensation; room-temperature superconductivity }

\author{M. de Llano}{
  address={Texas Center for Superconductivity, University of Houston,
Houston, TX 77204, USA and Instituto de Investigaciones en Materiales,
Universidad Nacional Aut\'{o}noma de M\'{e}xico, 04510 M\'{e}xico, DF, Mexico}}

\author{M. Grether}{ address={Facultad de Ciencias, UNAM, 04510 M\'{e}xico, DF, Mexico }}

\begin{abstract}
 By recognizing the vital importance of two-hole Cooper pairs (CPs) in
addition to the usual two-electron ones in a strongly-interacting
many-electron system, the concept of CPs was re-examined with striking
conclusions: namely, they are gapped and linearly-dispersive resonances with
a finite lifetime---but provided the ideal-gas Fermi sea is replaced by a
BCS-correlated unperturbed ground-state ``sea.'' Based on this,
Bose-Einstein condensation (BEC) theory has been ge\-ne\-ra\-lized to include not
boson-boson interactions (also neglected in BCS theory) but rather
boson-fermion (BF) interaction vertices reminiscent of the Fr\"{o}hlich
electron-phonon interaction in me\-tals. Instead of phonons, the bosons in the
generalized BEC (GBEC) theory are now \textit{both} particle and hole CPs.
Unlike BCS theory, the GBEC model is \textit{not} a mean-field theory
restricted to weak-coupling as it can be diagonalized exactly. It reproduces the BCS condensation energy exactly for any coupling, and each kind of CP is responsible for only \textit{half} the condensation
energy. The GBEC theory reduces to all the old known statistical theories as
special cases---including the so-called ``BCS-Bose crossover'' picture which
in turn generalizes BCS theory by \textit{not }assuming that the interelectronic
chemical potential equals the Fermi energy. Indeed, a BCS condensate is 
\textit{precisely} the weak-coupling limit of a GBE condensate with equal
numbers of both types of CPs. With feasible Cooper/BCS model interelectonic
interaction parameter values, and even without BF interactions, the GBEC
theory yields transition temperatures [including room-temperature
superconductivity (RTSC)] substantially higher than the BCS ceiling of
around 45K, without relying on non-phonon dynamics involving excitons,
plasmons, magnons or otherwise purely-electronic mechanisms. The results are
expected to shed light on the experimental search for RTSC.
\end{abstract}

\maketitle


\section{INTRODUCTION}

Boson-fermion (BF) models of superconductivity (SC) as a Bose-Einstein
condensation (BEC) go back to the mid-1950's \cite{Blatt}-\cite{BF2},
pre-dating even the BCS-Bogoliubov theory \cite{bcs}-\cite{bts}. Although
BCS theory only contemplates the presence of ``Cooper correlations'' of
single-particle states, BF models \cite{Blatt}-\cite{BF2},\cite%
{BF3}-\cite{Frontiers} posit the existence of actual bosonic CPs. In spite of the
central role played by CPs in both low- and high-$T_{c}$ superconductivity,
however, they are poorly understood. The fundamental drawback of early \cite%
{Blatt}-\cite{BF2} BF models, which took two-electron (2e) bosons as
analogous to diatomic molecules in a classical atom-molecule gas mixture, is
the notorious absence of an electron energy gap $\Delta (T)$. ``Gapless''
models are useful in locating transition temperatures if approached from
above, i.e., $T>T_{c}$. Even so, we are not aware of any calculations with
the early BF models attempting to reproduce any empirical $T_{c}$ values.
The gap first began to appear in later BF models \cite{BF3}-\cite{Frontiers}%
. With two \cite{BF7a}\cite{PLA2}\ exceptions, however, all BF models
neglect the effect of \textit{hole }CPs included on an equal footing with
electron CPs to give\ a GBEC theory consisting of \textit{both} bosonic CP
species coexisting with unpaired electrons in a \textit{ternary }BF model.
Although magnetic-flux-quantization measurements have established the
presence of \textit{pair }charge carriers in both conventional \cite%
{classical}-\cite{classical2}\ as well as cuprate \cite{cuprates}\
superconductors, no experiment has yet been done to our knowledge \cite%
{Gough} that distinguishes between electron and hole CPs.

The ``ordinary'' CP problem \cite{Coo}\ for two distinct interfermion
interactions (the $\delta $-well \cite{PRB2000}\cite{PhysicaC} or the
Cooper/ BCS model \cite{bcs}\cite{Coo}\ interactions) neglects the effect of
two-hole (2h) CPs treated on an equal footing with 2e-CPs---as Green's
functions \cite{FW}, on the other hand,\ can naturally ensure. However, a
crucial confirmed result \cite{PLA2}\ is that the BCS condensate is a very
particular BE condensate with \textit{equal numbers }of 2e- and 2h\textit{-}%
CPs, each contributing to one-half the condensation energy \cite{PC05}. This
was already evident, though not fully appreciated, from the perfect
symmetry\ about $\epsilon =\mu $, the electron chemical potential, of the
well-known Bogoliubov \cite{Bog} $v^{2}(\epsilon )$\ and $u^{2}(\epsilon )$
coefficients, where $\epsilon \equiv \hbar ^{2}k^{2}/2m$ is the electron
energy and $m$ its effective mass.\ The GBEC theory (also appropriately
viewed elsewhere\ as a ``complete boson-fermion model,'' or CBFM)
``unifies'' \cite{CMT02}\ both BCS and ordinary BEC theories as special
cases, and predicts substantially higher $T_{c}$'s than BCS theory with the
same Cooper/BCS model interaction that mimics electron-phonon dynamics. 

\section{SIGNIFICANCE OF THE COOPER INSTABILITY}

A Bethe-Salpeter (BS) many-body equation (in the ladder approximation)\
treating both 2e and 2h pairs on an equal footing reveals that, while the CP
problem [based on an ideal Fermi gas (IFG) ground state (the usual ``Fermi
sea'')] does \textit{not} possess energy solutions with a real part, it does
so when the IFG ground state is replaced by the BCS one. This is equivalent
to starting from an unperturbed Hamiltonian that is the BCS ground state
instead of the pure-kinetic-energy operator corresponding to the IFG. The
remaining Hamiltonian terms are then assumed amenable to a perturbation
treatment. As a result: i) CPs based not on the IFG-sea but on the BCS
ground state survive through a \textit{nontrivial} solution as
``generalized'' or ``moving''\ CPs which are \textit{positive} energy
resonances with an imaginary energy term leading to finite-lifetime effects;
ii) as in the ``ordinary'' CP problem, their dispersion relation in leading
order in the total (or center-of-mass) momentum (CMM) $\hbar \mathbf{K}\equiv
\hbar (\mathbf{k}_{1}+\mathbf{k}_{2})$ is also \textit{linear }(originally reported
without proof in Ref. \cite{Schrieffer}\textit{,} p.33)%
rather than the quadratic $\hbar ^{2}K^{2}/2(2m)$ of a composite boson
(e.g., a deuteron) of mass $2m$\ moving not in the Fermi sea but in vacuum;
and iii) this latter ``moving CP'' solution, though often confused with it,
is physically \textit{distinct }from another more common \textit{trivial}
solution sometimes called the Anderson-Bogoliubov-Higgs (ABH) \cite{Bog58}(%
\cite{bts} p. 44), \cite{ABH}\cite{Higgs} collective excitation. The ABH
mode is also linear in leading order and goes over into the IFG ordinary
sound mode in zero coupling. All this occurs in 1D \cite{FortesdeLlano05},
2D \cite{ANFdeLl} as well as in the 3D study outlined earlier in Ref. \cite%
{Honolulu}. In this section we focus on 2D because of its interest \cite%
{Brandow}\ for quasi-2D high-$T_{c}$ cuprate superconductors. In general,
the results will be crucial for BEC scenarios employing BF models of
superconductivity, not only \textit{in} \textit{exactly 2D} as with the
Berezinskii-Kosterlitz-Thouless (BKT) \cite{BKT}\cite{KT} transition, but
also down to ($1+\epsilon $)D which characterize the quasi-1D
organo-metallic (Bechgaard salt) \cite{organometallics}-\cite{jerome2}, and
most recently multi-walled carbon nanotube \cite{Peter}, SCs.

If $\mathbf{K\equiv k}_{1}+\mathbf{k}_{2}$ is the CMM and $\mathbf{k\equiv 
{\frac12}(k}_{1}-\mathbf{k}_{2}\mathbf{)}$ the relative momentum wavevectors of the
2e bound state, and $\mathcal{E}_{K}$ $\equiv E_{1}+E_{2}$ is its energy
with $E_{1}$ and $E_{2}$ the energies of electrons $1$ and $2$,\ one
uses the bare one-fermion Green's function $G_{0}\left( \mathbf{k}_{1}\equiv 
\mathbf{K}/2+\mathbf{k},E_{1}\equiv \mathcal{E}_{K}/2+E\right) $\ for
particle $1$, and similarly for particle $2$, where $E\equiv 
{\frac12}(E_{1}-E_{2})$. The solution of the \textit{complete }BS equation
based on the IFG unperturbed state with \textit{both }2e- and 2h-CPs
included, and formed via the Cooper/BCS model interaction, is 
\begin{equation}
\mathcal{E}_{0}=\pm i2\hbar \omega _{D}/\sqrt{e^{2/\lambda }-1}  \label{18}
\end{equation}
where $\lambda \equiv VN(E_{F})$ with $N(E_{F})$ the electronic density of
states (DOS) for one spin, while $V$ is a positive coupling constant and ${%
\hbar \omega _{D}}$\ an energy cutoff, both defined below in (\ref%
{BCSint}). As the CP energy is pure-imaginary there is an obvious
instability of the CP problem when both type pairs are included. This result
was originally derived in Refs. \cite{bts} p. 44 and \cite{AGD}; also, it was
guessed in Ref. \cite{Schrieffer}\ p. 167 without explicit mention of
hole-pairing.\ It\ contrasts sharply with the familiar solution \cite{Coo}\
for 2e-CPs only, namely%
\begin{equation}
\mathcal{E}_{0}=-{2\hbar \omega _{D}/(}e^{2/\lambda }-1)\mathrel{\mathop{%
\longrightarrow}\limits_{\lambda \rightarrow 0}}-2\hbar \omega
_{D}e^{-2/\lambda }.  \label{Cooper}
\end{equation}
The first expression is exact in 2D and a very good approximation otherwise
if $\hbar \omega _{D}\ll E_{F}$, where $\omega _{D}$ is the Debye frequency.
The sometimes misnamed ``Cooper instability'' (\ref{Cooper}) merely represents a
negative-energy, stationary-state (i.e., infinite-lifetime) bound pair. We
suggest, however, that unlike the apparent negative-but-real-$\mathcal{E}_{0}
$\ ``instability'' (\ref{Cooper}) the genuine Cooper instability\ is really (%
\ref{18}) so that the original CP picture \textit{is meaningless if 2e- and
2h-CPs are treated on an equal footing}, as consistency demands, since it
leads to a purely-imaginary eigenvalue $\mathcal{E}_{0}$.

However, a BS treatment of pairs referred not to the IFG sea but to a
BCS-correlated ground state ``sea''\ \textit{vindicates the CP concept }in
terms of a new nontrivial solution. This is tantamount to starting not from
the IFG unperturbed Hamiltonian but from the BCS one. Its physical
justification is reinforced through the recovery of three expected results:
a) the (trivial) ABH sound mode solution; b) the BCS $T=0$ gap equation;\
and c) \textit{finite}-lifetime effects of a ``moving-CP'' nontrivial
solution in either 2D \cite{ANFdeLl} or 3D \cite{Honolulu}.\ Thus, the IFG
Green function $G_{0}\left(\mathbf{k}_{1},E_{1}\right) $ is replaced by the
BCS one, say, \textbf{G}$_{0}(\mathbf{k}_{1},E_{1})$ that now refers to an
energy $E_{1}\equiv E_{k_{1}}\equiv \sqrt{\xi _{k_{1}}{}^{2}+\Delta
^{2}}$ with $\xi _{k}\equiv \hbar ^{2}k^{2}/2m-E_{F}\ $and $\Delta $ the $T=0
$ fermionic gap. It also contains the Bogoliubov functions \cite{Bog} $%
v_{k}^{2}\equiv \frac{1}{2}(1-\xi _{k}/E_{{k}})$ and $u_{k}^{2}\equiv
1-v_{k}^{2}$.\ There are then \textit{two }solutions to the BS equations. A
trivial solution is the ABH energy eigenvalue $\mathcal{E}_{K}$, which when
Taylor-expanded about $K=0$ gives for small $\lambda $ in 2D%
\begin{equation}
\mathcal{E}_{K}=\frac{\hbar v_{F}}{\sqrt{2}}K+O(K^{2})+o(\lambda ),
\label{22}
\end{equation}%
where $o(\lambda )$ denote interfermion interaction correction terms that
vanish as $\lambda \rightarrow 0$.\ Note that the leading term is just the
ordinary sound mode in an IFG with sound speed $v_{F}/\sqrt{d}$ in $d$
dimensions, as determined straightforwardly from standard thermodynamic
formulae. Secondly, there is a nontrivial \textit{moving CP}\ solution of
the BCS-correlated-sea-based BS treatment, which is \textit{entirely new}
and leads to the pair energy $\mathcal{E}_{K}$ which in 2D is \cite{ANFdeLl}%
\begin{equation}
\pm \mathcal{E}_{K}=2\Delta +\frac{\lambda }{2\pi }\hbar v_{F}K+\frac{1}{9}%
\frac{\hbar v_{F}}{k_{D}}e^{1/\lambda }K^{2}-i\left[ \frac{\lambda }{\pi }%
\hbar v_{F}K+\frac{1}{12}\frac{\hbar v_{F}}{k_{D}}e^{1/\lambda }K^{2}\right]
+O(K^{3}).  \label{linquadmCP}
\end{equation}%
Here, the upper and lower signs refer to 2e- and 2h-CPs, respectively, and $%
k_{D}\equiv \omega _{D}/v_{F}$ with $\omega _{D}$ the Debye frequency. A
linear dispersion in leading order again appears, but now associated with
the bosonic moving CP. From (\ref{linquadmCP}) the \textit{positive}-energy
2p-CP resonance has an energy width $\Gamma _{K}$ and a lifetime $\tau
_{K}\equiv \hbar /2\Gamma _{K}=\hbar /2\left[ (\lambda /\pi )\hbar
v_{F}K+(\hbar v_{F}/12k_{D})e^{1/\lambda }K^{2}\right] $ that diverges at $%
K=0$, falling to zero as $K$ increases. Thus, ``faster'' moving CPs are
shorter-lived and eventually break up, while ``non-moving'' $K=0$\ ones are
in infinite-lifetime stationary states.

\section{GENERALIZED BEC THEORY OF SUPERCONDUCTORS}

The GBEC theory \cite{BF7a,PLA2} is described in $d$ dimensions by the
Hamiltonian $H=H_{0}+H_{int}$. The unperturbed Hamiltonian $H_{0}$\
corresponds to a non-Fermi-liquid ``normal'' state which is an \textit{ideal 
}(i.e., noninteracting) ternary gas mixture of unpaired fermions and both
types of CPs, two-electron (2e) and two-hole (2h), namely%
\begin{equation}
H_{0}=\sum\limits_{\mathbf{k}_{1},s}\epsilon
_{_{k_{1}}}a_{k_{1},s}^{+}a_{k_{1},s}+\sum\limits_{\mathbf{K}}E_{+}(K)b_{%
\mathbf{K}}^{+}b_{\mathbf{K}}-\sum\limits_{\mathbf{K}}E_{-}(K)c_{\mathbf{K}%
}^{+}c_{\mathbf{K}}  \label{H0}
\end{equation}%
where as before $\mathbf{K\equiv k}_{\mathbf{1}}+\mathbf{k}_{\mathbf{2}}$ is
the CP CMM wavevector while $\epsilon _{k_{1}}\equiv \hbar ^{2}k_{1}^{2}/2m$%
, e.g., are the single-electron, and $E_{\pm }(K)$\ the 2e-/2h-CP \textit{%
phenomenological, }energies.\ Here $a_{\mathbf{k},s}^{+}$ ($a_{\mathbf{k},s}$%
) are creation (annihilation) operators for fermions. Similarly $b_{%
\mathbf{K}}^{+}$ ($b_{\mathbf{K}}$) and $c_{\mathbf{K}}^{+}$ ($c_{\mathbf{K}}
$) are such for 2e- and 2h-CP bosons, respectively---although we do not attempt to
construct them starting from the fermion operators. Two-hole CPs are
postulated to be \textit{distinct }and\textit{\ kinematically independent }%
from 2e-CPs, all of which provides a \textit{ternary} BF gas mixture. This
postulate is firmly grounded on the experiments \cite{classical}-\cite%
{cuprates} cited before. The operator $H_{0}$ then has diagonal form and its 
\textit{exact} eigenstates can be numerated by the sets of occupation
numbers $\{...n_{\mathbf{k},s}...N_{\mathbf{K}}...M_{\mathbf{K}}...\}$. The
occupation numbers $n_{\mathbf{k},s}$ of one-fermion states each take on
only the two values $0$ and $1$, while those of the one-boson momentum-$%
\mathbf{K}$ states of 2p-CPs $N_{\mathbf{K}}$, and of 2h-CPs $M_{\mathbf{K}}$%
, take on all values $0,1,2,\cdots \infty $.

The exact eigenstates of the Hamiltonian $H_{0}$ are then 
\begin{equation}
\mid ...n_{\mathbf{k},s}...N_{\mathbf{K}}...M_{\mathbf{K}}...\rangle
=\prod\limits_{\mathbf{k},s}(a_{\mathbf{k},s}^{+})^{n_{\mathbf{k}%
,s}}\prod\limits_{\mathbf{K}}\frac{1}{\sqrt{N_{\mathbf{K}}!}}\left( b_{%
\mathbf{K}}^{+}\right) ^{N_{\mathbf{K}}}\prod\limits_{\mathbf{K}}\frac{1}{%
\sqrt{M_{\mathbf{K}}!}}\left( c_{\mathbf{K}}^{+}\right) ^{M_{\mathbf{K}%
}}\mid O\rangle   \label{fihoprimprim}
\end{equation}%
where $\mid O\rangle $ is the vacuum state for fermions and simultaneously
for 2e-CP and 2h-CP {creation and annihilation operators. Specifically, }%
\begin{equation}
a_{\mathbf{k},s}\mid O\rangle \equiv b_{\mathbf{K}}\mid O\rangle \equiv c_{%
\mathbf{K}}\mid O\rangle \equiv 0.
\end{equation}%
If $N_{op}$ is the number operator of the total number of electrons, charge
conservation implies that

\begin{equation}
N_{op}=\sum\limits_{\mathbf{k}_{1},s_{1}}a_{\mathbf{k}_{1},s_{_{1}}}^{+}a_{%
\mathbf{k}_{1},s_{_{1}}}+2\sum\limits_{\mathbf{K}}b_{\mathbf{K}}^{+}b_{%
\mathbf{K}}-2\sum\limits_{\mathbf{K}}c_{\mathbf{K}}^{+}c_{\mathbf{K}}.
\end{equation}%
If $\mu $ is their chemical potential, the exact eigenvalues of $H_{0}-\mu
N_{op}$ are%
\begin{eqnarray}
E_{...n_{\mathbf{k},s}...N_{\mathbf{K}}...M_{\mathbf{K}}...}
&=&[E_{+}(0)-2\mu ]N_{0}+[2\mu -E_{-}(0)]M_{0}+\sum\limits_{\mathbf{k}
,s}(\epsilon _{k}-\mu )n_{\mathbf{k},s}  \nonumber \\
&&+\sum\limits_{\mathbf{K}\neq 0}[E_{+}(K)-2\mu ]N_{\mathbf{K}}+\sum\limits_{
\mathbf{K}\neq 0}[2\mu -E_{-}(K)]M_{\mathbf{K}}.
\end{eqnarray}

The interaction Hamiltonian $H_{int}$ part of $H=H_{0}+H_{int}$\ consists of
four distinct BF interaction vertices each with two-fermion/one-boson
creation or annihilation operators, depicting how unpaired electrons
(subindex +) [or holes (subindex $-$)] combine to form the 2e- (and 2h-) CPs
assumed in the $d$-dimensional system of size $L$, namely 
\begin{eqnarray}
H_{int}&=&L^{-d/2}\sum\limits_{\mathbf{k},\mathbf{K}}f_{+}(k)\{a_{\mathbf{k}+%
\frac{1}{2}\mathbf{K},\uparrow }^{+}a_{-\mathbf{k}+\frac{1}{2}\mathbf{K}%
,\downarrow }^{+}b_{\mathbf{K}}+a_{-\mathbf{k}+\frac{1}{2}%
\mathbf{K},\downarrow }a_{\mathbf{k}+\frac{1}{2}\mathbf{K},\uparrow }b_{%
\mathbf{K}}^{+}\} \nonumber \\
&+&L^{-d/2}\sum\limits_{\mathbf{k},\mathbf{K}}f_{-}(k)\{a_{\mathbf{k}+\frac{1}{%
2}\mathbf{K},\uparrow }^{+}a_{-\mathbf{k}+\frac{1}{2}\mathbf{K},\downarrow
}^{+}c_{\mathbf{K}}^{+}+a_{-\mathbf{k}+\frac{1}{2}\mathbf{K}%
,\downarrow }a_{\mathbf{k}+\frac{1}{2}\mathbf{K},\uparrow }c_{\mathbf{K}}\}
\label{Hint}
\end{eqnarray}
where $\mathbf{k}\equiv \frac{1}{2}\mathbf{(k}_{\mathbf{1}}-\mathbf{k}_{%
\mathbf{2}})$ is the relative wavevector of a CP. The energy form factors $%
f_{\pm }(k)$ in (\ref{Hint}) are\ essentially the Fourier transforms of the
2e- and 2h-CP intrinsic wavefunctions, respectively, in the relative
coordinate of the two fermions. The GBEC Hamiltonian $H=H_{0}+H_{int}$ is
very different from the well-known BCS Hamiltonian%
\begin{equation}
H^{BCS}\equiv H_{0}^{BCS}+H_{int}^{BCS}=\sum\limits_{\mathbf{%
k}_{1},s_{1}}\epsilon_{k_{1}}a_{k_{1},s_{_{1}}}^{+}a_{k_{1},s_{_{1}}}+\sum\limits_{\mathbf{k}_{1},%
\mathbf{l}_{1}}V_{\mathbf{k}_{1}\mathbf{,l}_{1}}a_{\mathbf{k}_{1}\uparrow
}^{+}a_{-\mathbf{k}_{1}\downarrow }^{+}a_{-\mathbf{l}_{1}\downarrow }a_{%
\mathbf{l}_{1}\uparrow }  \label{BCSH}
\end{equation}%
with, say, the Cooper/BCS model interaction, with $V>0,$%
\begin{equation}
V_{\mathbf{k}_{1}\mathbf{,l}_{1}}=\left\{ 
\begin{array}{cc}
-V/L^{d} & {\ \ if }\, \,\, \, \, \,  \mu -\hbar \omega _{D}<\epsilon _{k_{1}},%
\epsilon _{l_{1}}<\mu +\hbar \omega _{D} \\ 
0 & {otherwise.}
\end{array}
\right.   \label{BCSint}
\end{equation}

In Refs. \cite{BF7a}\cite{PLA2} the energy form factors $f_{\pm }(k)$ for
the GBEC theory are taken as%
\begin{equation}
f_{\pm }(\epsilon )=\left\{ 
\begin{array}{cc}
f & {\ \ if }\, \, \, \, \, \, {\frac12}[E_{\pm }(0)-\delta \varepsilon ]<\epsilon <{\frac12}
[E_{\pm }(0)+\delta \varepsilon ] \\ 
0 & {otherwise}
\end{array}
\right. 
\end{equation}
in order for it to reduce properly to BCS theory with (\ref{BCSint}) in the
limit to be explained below, provided one identifies $f$ and $\delta
\varepsilon $ with $\sqrt{2V\hbar \omega _{D}}$ and $\hbar \omega _{D}$,
respectively. One then introduces the quantities $E_{f}$ and $\delta
\varepsilon $ as new phenomenological dynamical energy parameters
(in addition to the positive BF vertex coupling parameter $f$) that replace
the previous such $E_{\pm }(0)$, through the definitions 
\begin{equation}
E_{f}\equiv \frac{1}{4}[E_{+}(0)+E_{-}(0)]{ \ \ and \ \ }\delta
\varepsilon \,\equiv \,\frac{1}{2}[E_{+}(0)-E_{-}(0)]{  \ \ }%
\Rightarrow {  \ \ }E_{\pm }(0)=2E_{f}\pm \delta \varepsilon 
\label{27}
\end{equation}
where $E_{\pm }(0)$ are the (empirically \textit{un}known) zero-CMM energies
of the 2e- and 2h-CPs, respectively. The quantity\ $E_{f}$ is available as a
possibly convenient energy scale. It is not to be confused with the Fermi
energy $E_{F}=\frac{1}{2}mv_{F}^{2}\equiv \hbar ^{2}k_{F}^{2}/2m\equiv
k_{B}T_{F}$ where $T_{F}$\ is the Fermi temperature. If $n\equiv N/L^{d}$ is
the total number-density of charge-carrier electrons, then $n=k_{F}^{2}/2\pi 
$ in 2D and $=k_{F}^{3}/3\pi ^{2}$ in 3D.\ Thus, the Fermi energy $E_{F}$
equals $\pi \hbar ^{2}n/m$ in 2D and $(\hbar ^{2}/2m)(3\pi ^{2}n)^{2/3}$ in
3D, while $E_{f}$ equals $\pi \hbar ^{2}n_{f}/m$ in 2D and $(\hbar
^{2}/2m)(3\pi ^{2}n_{f})^{2/3}$ in 3D, i.e., is the same as $E_{F}$\ with $n$
replaced by $n_{f}$ which will serve as a convenient density scale. The
quantities $E_{f}$ and $E_{F}$ coincide \textit{only }when perfect 2e/2h-CP
symmetry holds, i.e., when $n=n_{f}$.

The interaction Hamiltonian (\ref{Hint}) can be further simplified by
dropping all $\mathbf{K}\neq 0$ terms. This is also done in BCS theory but
in the \textit{full }BCS Hamiltonian (\ref{BCSH}). However, in the GBEC
theory these terms are retained in (\ref{H0}) so that $H_{0}$
describes a \textit{non-Fermi-liquid normal state}. Following Bogoliubov %
\cite{Bog47}, the boson operators $b_{0}^{+},b_{0}$\ and $c_{0}^{+},c_{0}$
remaining in (\ref{Hint}) are then replaced by the ``c-numbers'' $\sqrt{%
N_{0}(T)}$\ and $\sqrt{M_{0}(T)}$, respectively, where $N_{0}$ and $M_{0}$
are the number of $K=0$ 2e- and 2h-CPs.\ The full GBEC Hamiltonian $H$ thus
becomes bilinear in\ the fermion operators $a_{\mathbf{k},s}^{+}$, $a_{%
\mathbf{k},s}$\ and is diagonalizable via a Bogoliubov transformation 
\textit{exactly }without assuming weak coupling as in BCS theory.\ One
constructs the grand potential $\Omega $\ for the full GBEC Hamiltonian as%
\begin{equation}
\Omega (T,L^{d},\mu ,N_{0},M_{0})=-k_{B}T\ln \left[ Tr \ \exp \ \{-\beta
(H-\mu N_{op})\}\right]   \label{28}
\end{equation}%
where ``$Tr$'' stands for ``trace'' and $\beta \equiv 1/k_{B}T.$ Minimizing
with respect to $N_{0}$ (the number of zero-CMM 2e-CPs) and $M_{0}$ (the
same for 2h-CPs), while simultaneously fixing the total number $N$ of
electrons via the electron chemical potential $\mu $, determines an \textit{%
equilibrium state} of the system with volume $L^{d}$ and temperature $T$.
One thus requires that 
\begin{equation}
\frac{\partial \Omega }{\partial N_{0}}=0,\ \ \ \ \ \ \ \ \ \ \ \
\frac{\partial \Omega }{\partial M_{0}}={
\ }0{ \ \ \ \ \ \ \ \ \ \ \ \ and \ \ \ \ \ \ \ \ \ \ \ \ }\frac{%
\partial \Omega }{\partial \mu }{ \ }={ \ }-N.  \label{36}
\end{equation}%
Here $N$ evidently includes both paired and unpaired CP electrons. Some
algebra then leads to the three coupled integral Eqs. (7)-(9) of Ref. \cite%
{BF7a}. The relation between the fermion spectrum $E(\epsilon )$ and fermion
energy gap $\Delta (\epsilon )$ turns out to be of the BCS form%
\begin{equation}
E(\epsilon )=\sqrt{(\epsilon -\mu )^{2}+\Delta ^{2}(\epsilon )}{ \ \ \
\ but \, where \ \ \ \ }\Delta (\epsilon )\equiv \sqrt{n_{0}}f_{+}(\epsilon )+%
\sqrt{m_{0}}f_{-}(\epsilon ).  \label{34}
\end{equation}%
This last expression for the gap $\Delta (\epsilon )$ implies a simple $T$%
-dependence rooted in the 2e-CP $n_{0}(T)\equiv N_{0}(T)/L^{d}$ and 2h-CP $%
m_{0}(T)\equiv M_{0}(T)/L^{d}$ number densities of BE-condensed bosons, i.e.,%
\begin{equation}
\Delta (T)=\sqrt{n_{0}(T)}f_{+}(\epsilon )+\sqrt{m_{0}(T)}f_{-}(\epsilon ).
\label{35}
\end{equation}%
This generalizes the relation between BCS and BEC order parameters first
reported in Ref. \cite{BF3}. Self-consistent (at worst, numerical) solution
of the aforementioned \textit{three coupled equations}\textbf{\ }then yields
the three thermodynamic variables of the GBEC theory%
\begin{equation}
\ \ n_{0}(T,n,\mu ),\;\ \ \ m_{0}(T,n,\mu )\ \ \ \ {and}\ \ \ \ \mu
(T,n).  \label{43}
\end{equation}
\vskip -0.0truecm
\begin{figure}
\includegraphics[height=0.5\textheight]{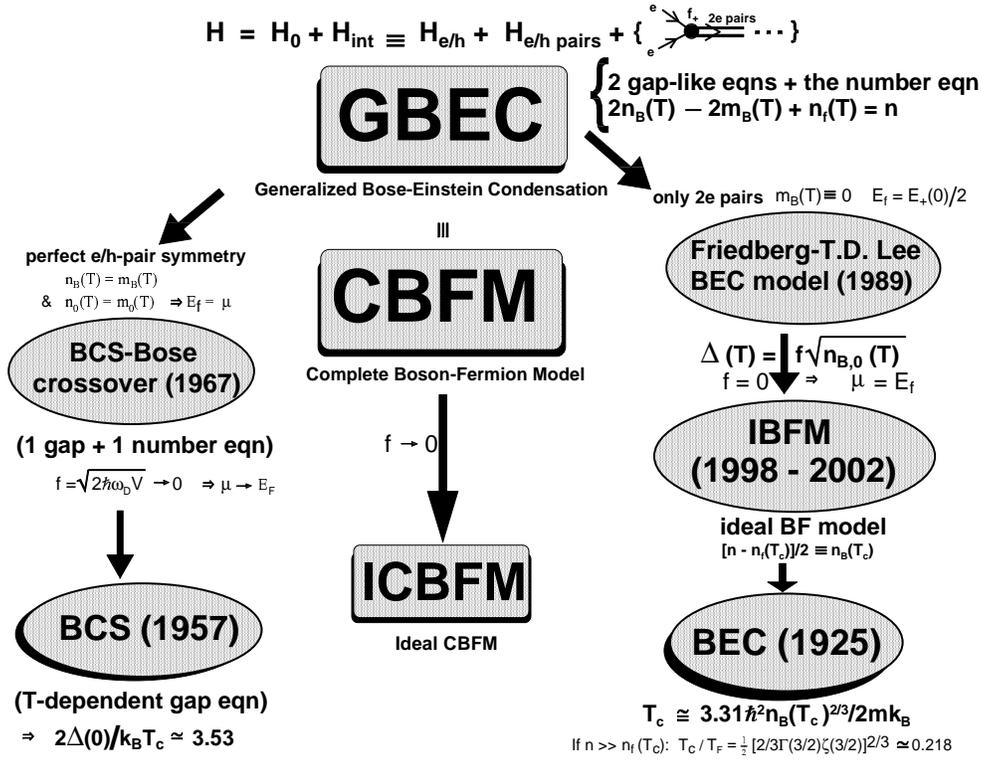}
\caption{Flowchart outlining conditions under which the GBEC [or ``complete
boson-fermion model'' (CBFM)] theory reduces to all five statistical
theories of superconductors (ovals). Symbols are explained throughout text.}
\end{figure}

Most significantly, the three GBEC theory equations contain the key
equations of \textit{five }different statistical theories as special cases;
for a detailed review see Ref. \cite{Frontiers}. Figure 1 illustrates this
in a flowchart. Perfect 2e/2h CP symmetry signifies equal number of 2e- and
2h-CPs, i.e., $n_{B}(T)=m_{B}(T)$ \textit{as well as }$n_{0}(T)=m_{0}(T).$
With (\ref{27}) Eqs. (\ref{40}) and (\ref{41}) below imply that $E_{f}$ coincides with $\mu $, and the
GBEC theory then reduces to:

\textbf{i)} the gap and number equations of the \textit{BCS-Bose crossover
picture} for the BCS model interaction---if the BCS parameters $V$ and Debye
energy $\hbar \omega _{D}$ are identified with the BF interaction
Hamiltonian $H_{int}$ parameters $f^{2}/2\delta \varepsilon $ and $\delta
\varepsilon $, respectively. The crossover picture for unknowns $\Delta (T)$
and $\mu (T)$ is now supplemented by the key expression relating BCS with
BEC precisely, namely 
\begin{equation}
\Delta (T)=f\sqrt{n_{0}(T)}=f\sqrt{m_{0}(T)}.  \label{sqrt}
\end{equation}
The crossover picture is associated with many authors beginning in 1967 with
Friedel and coauthors \cite{Friedel} and then given mayor impetus by Eagles %
\cite{Eagles} who in turn introduced the BEC mechanism into the picture; for
reviews see Refs. \cite{BCS-Bose}\cite{LevinPhysReps}. However, it is widely
unrecognized to be a very modest improvement, at least for the Cooper/BCS
model interaction, over BCS theory \textit{per se}\ since, e.g., an
unphysically large $\lambda $ of about 8 is required to bring $\mu
(T_{c})/E_{F}$ in 2D down from 1.00 to 0.998; indeed, $T_{c}$-values differ very
slightly \cite{PC05} in 2D between the crossover and BCS theories all the way up
to $\lambda \sim 50$ when the Fermi surface originally pinned at $\mu $\
disappears by becoming negative so that the model interaction breaks down.
In fact, room-temperature superconductivity is predicted \cite{PC05}\ by BCS
theory in 2D for a material with $T_{F}=10^{3}$ $K$,\ but only for $\lambda $
values somewhat larger than $10$; these are still too unphysical\ as they
exceed the Migdal \cite{Migdal}\ threshold of $\lambda >{\frac12}$
for ionic-lattice instability. If one imposes that $\mu (T_{c})=E_{F}$
exactly, as follows from the number equation for weak BF coupling $f$,\ the
crossover picture is well-known to reduce to:

\textbf{ii)} \textit{ordinary BCS theory }which is characterized by a 
\textit{single} equation, the gap equation for any $T$. Thus, \textit{the
BCS condensate is precisely a BE condensate} whenever both $n_{B}(T)=m_{B}(T)
$ and $n_{0}(T)=m_{0}(T)$ \textit{and }the BF coupling $f$ is small. The condensation energies of the GBEC and BCS theories coincide exactly for all coupling.

On the other hand, for no 2h-CPs present the GBEC theory reduces \cite{BF7a}%
\ also to:

\textbf{iii)} the \textit{BEC BF model} in 3D of Friedberg and T.D. Lee \cite%
{BF5,BF6} characterized by the relation $\Delta (T)=f\sqrt{n_{0}(T)}$ first
reported in Ref. \cite{BF3}; but lacking 2h-CPs this model cannot be fully
related to BCS theory. When $f=0$ it reduces\ to:

\textbf{iv)} the \textit{ideal BF model} (IBFM) of Refs. \cite{BF9,BF10}
that predicts nonzero 2e-CP BEC $T_{c}$'s even in 2D. The ``gapless'' IBFM\
cannot describe the superconducting phase. But considered as a model for the 
\textit{normal state} it should provide feasible $T_{c}$'s as singularities
within a BE scenario that are approached from \textit{above }$T_{c}$, and
this is indeed \cite{BF10} the case. Finally, in the limit of no unpaired
electrons this model in 3D\ reduces to:

\textbf{v)} the familiar $T_{c}$-formula of ordinary BEC in 3D but where the
boson number-density is temperature dependent.

The vastly more general GBEC theory has been applied in both 2D and 3D and
gives sizeable enhancements in $T_{c}$'s over BCS theory for moderate
departures from perfect 2e/2h-pair symmetry. This is attained for the 
\textit{same }Cooper/BCS interaction model (coupling strength $\lambda $ and
cutoff $\hbar \omega _{D}$) parameter values often used in conventional SCs.
The three coupled equations of the GBEC theory\ that determine the $d$%
-dimensional BE-condensate number-densities $n_{0}(T)$ and $m_{0}(T)$ of 2e-
and 2h-CPs, respectively, as well as the electron chemical potential $\mu (T)
$, were first solved numerically \cite{PLA2} around the BCS point of the $%
T_{c}$\ vs. $n$\ phase diagram. At $n/n_{f}=1$ one has perfect 2e/2h-CP
symmetry; the plain $n_{f}$ can be seen to be the number density $n_{f}(T)$\
of unpaired but BCS-correlated electrons when $\Delta =0$ and $T=0,$
whenever $\mu =E_{f}$. In general
\begin{equation}
n_{f}(T)\equiv \int\limits_{0}^{\;\infty }d\epsilon N(\epsilon )[1-\frac{%
\epsilon -\mu }{E(\epsilon )}\tanh{\frac12}
\beta E(\epsilon )]  \label{unpaired}
\end{equation}
with $E(\epsilon )$ defined in (\ref{34}).\ This expression is \textit{%
precisely} the BCS expression for the electron number density 
\begin{equation}
n=\sum\limits_{\mathbf{k},s}{v}_{k}^{2}(T)  \label{BCSnumeq}
\end{equation}%
where $v_{k}(T)$ is the temperature-dependent Bogoliubov function. Alongside
two gap-like equations involving $n_{0}(T)$ and $m_{0}(T)$, the third, or
``complete'' number, equation explicitly reads%
\begin{equation}
n_{f}(T)+2n_{0}(T)+2n_{B+}(T)-2m_{0}(T)-2m_{B+}(T)=n  \label{57'}
\end{equation}%
with $m_{B+}(T)$, e.g., being precisely the number of ``pre-formed'' $K>0$\
2h-CPs, and $n_{B+}(T)$\ the same for 2e-CPs.\ Besides the \textit{normal }%
phase consisting of the ideal BF ternary gas described by $H_{0}$, three
different stable BEC phases emerge: two pure phases consisting of a 2e-CP
BEC and a 2h-CP BEC, as well as a mixed phase consisting of both types of
BECs in varying proportions. For a half-and-half mixed phase all the boson
number-density terms in (\ref{57'})\ cancel out and the BCS number equation $%
n_{f}(T)=n$\ is recovered.

We shall focus on the \textit{linear }dispersion\ that occurs in leading
order in $K$ for ``ordinary'' CPs in a Fermi sea as well as for
``generalized'' CPs in a BCS-correlated state.\ For the latter, the boson
excitation energy $\varepsilon $ to be used has a leading term in the
many-body Bethe-Salpeter (BS) CP dispersion relation given by $\varepsilon
\simeq (\lambda /4)\hbar v_{F}K$ in 3D \cite{Honolulu}, as part of an
expansion similar to (\ref{linquadmCP})\ in 2D. As before, $\lambda \equiv
VN(E_{F})$ where $N(E_{F})$ is the electron DOS (for one spin) at the Fermi
surface. Note that $\varepsilon $ is no longer the quadratic $%
\hbar ^{2}K^{2}/2(2m)$ often assumed \cite{Blatt}-\cite{BF2}, \cite{BF5}-%
\cite{PLA2}, \cite{LevinPhysReps} and associated with a composite boson of
mass $2m$ moving not in the Fermi sea but in vacuum.

\section{CRITICAL TEMPERATURES IN 3D: GBEC WITH $f$ = 0 }

Fully equivalent to a ``complete boson-fermion model'' (CBFM), the GBEC theory
with $f=0$ becomes an ``ideal boson-fermion model'' (ICBFM), see Fig. 1. The
ICBFM is completely\ described in $d$ dimensions by the Hamiltonian $H_{0}$
defined by (\ref{H0}). One can construct its associated grand potential as 
\begin{equation}
\Omega _{0}(T,L^{d},\mu ,N_{0},M_{0}){\ }=-k_{B}T\ln \left[ {Tr \ \exp%
}\{-\beta (H_{0}-\mu N_{op})\}\right] 
\end{equation}
with $N_{0}$ and $M_{0}$ as before the number of zero-CMM 2e- and 2h-CPs,
respectively. One gets
\begin{eqnarray}
&&\Omega _{0}(T,\,L^{d},\mu ,\,N_{0},\,M_{0})=[E_{+}(0)-2\mu
]N_{0}+[2\mu -E_{-}(0)]M_{0}  \nonumber \\
&-&2k_{B}TB_{d}L^{d}\int\limits_{0}^{\infty }k^{d-1}dk\ln [1+{\exp}%
\{-\beta (\epsilon _{k}-\mu )\}]  \nonumber \\
&+&k_{B}TB_{d}L^{d}\int\limits_{0^{+}}^{\infty }K^{d-1}dK\ln (1-{\exp}%
\{-\beta \lbrack E_{+}(K)-2\mu ]\})  \nonumber \\
&+&k_{B}TB_{d}L^{d}\int\limits_{0^{+}}^{\infty }K^{d-1}dK\ln (1-{\exp}
\{-\beta \lbrack 2\mu -E_{-}(K)]\})  \label{potvol}
\end{eqnarray}%
where $B_{d}=1/\pi $, $1/2\pi $ and $1/2\pi ^{2}$ for $d=1,$ $2$ and $3,$
respectively. An equilibrium thermodynamic state makes $\Omega_0 (T,L^{d},\mu
,N_{0},M_{0})$ stationary with respect to $N_{0}$ and to $M_{0}$ and
requires that the number density of electrons $n\equiv N/L^{d}$ remain
constant. Thus, one imposes that
\begin{equation}
\frac{\partial \Omega _{0}}{\partial N_{0}}=0,\ \ \ \ \ \ \ \ \ \frac{%
\partial \Omega _{0}}{\partial M_{0}}=0\ \ \ \ \ \ \ \ \ \ {and}\ \ \ \
\ \ \ \ \ \ -\frac{\partial \Omega _{0}}{\partial \mu }=N.
\label{3conditions}
\end{equation}
This leads to the three relations
\begin{equation}
\lbrack E_{+}(0)-2\mu ]=0,{ \ \ \ \ \ }[2\mu -E_{-}(0)]=0{ \ \ \ \ \ and \ \ \ \ \ }n_{f}(T)+2n_{B}(T)-2m_{B}(T)=n  \label{Emas}
\end{equation}
with the latter being the \textit{``number equation'' }that ensures charge
conservation in the ternary mixture. In the last relation 
\begin{eqnarray}
n_{B}(T) &\equiv &n_{0}(T)+B_{d}\int\limits_{0+}^{\infty }dKK^{d-1}[{\exp\{}\beta \lbrack E_{+}(0)-2\mu +\varepsilon _{K}]{\}}-1]^{-1}
\label{40} \\
m_{B}(T) &\equiv &m_{0}(T)+B_{d}\int\limits_{0+}^{\infty }dKK^{d-1}[{\exp}\{\beta \lbrack 2\mu -E_{-}(0)+\varepsilon _{K}]-1]^{-1}  \label{41}
\end{eqnarray}%
while 
\begin{equation}
n_{f}(T)\equiv 2B_{d}\int\limits_{0}^{\infty }dkk^{d-1}[{\exp}\{\beta
(\epsilon _{k}-\mu )\}+1]^{-1}.  \label{42}
\end{equation}%
This last expression can be interpreted as the number density of \textit{%
unpaired} electrons at any $T$ and $\epsilon _{k}=\hbar ^{2}k^{2}/2m$ is the
electron energy.

Consider the three equations (\ref{Emas}) assuming only that $%
E_{-}(0)<E_{+}(0)$. In 3D, the third equation of (\ref{Emas}) explicitly
becomes 
\begin{eqnarray}
n &\equiv &k_{F}^{3}/3\pi ^{2}=\pi ^{-2}\int\limits_{0}^{\infty }k^{2}dk[
{\exp}\{\beta (\epsilon _{k}-\mu )\}+1]^{-1}  \nonumber \\
&&+2n_{0}(T)+\pi ^{-2}\int\limits_{0}^{\infty }K^{2}dK({\exp}\{\beta
\lbrack E_{+}\mathbf{(}0\mathbf{)}-2\mu +\varepsilon _{K}]\}-1)^{-1}  \nonumber
\\
&&-2m_{0}(T)-\pi ^{-2}\int\limits_{0}^{\infty }K^{2}dK({\exp}\{\beta
\lbrack 2\mu -E_{-}(0)+\varepsilon _{K}]\}-1)^{-1}.  \label{sistema}
\end{eqnarray}%
Take first the limiting case $2\mu =E_{+}(0)$ when the first equation of (%
\ref{Emas}) is satisfied but not the second. From (\ref{3conditions}) this
implies that $M_{0}(T)$ must vanish for all $T$, but that $N_{0}(T)\neq 0$
at least for some $T$\ so that a pure 2e-CP BEC phase may occur below a
critical temperature $T_{c}$ (possibly zero) determined by $n_{0}(T_{c},n)=0$. After
substituting (\ref{27}) in (\ref{sistema}) with $\mu =E_{+}(0)/2$, thus
eliminating $E_{\pm }(0)$ in favor of $E_{f}\equiv (\hbar ^{2}/2m)(3\pi
^{2}n_{f})^{2/3}$and $\delta \varepsilon \equiv \hbar \omega _{D}$, we
obtain the dimensionless ``working number equation'' for the\textit{\ pure
2e-CP BEC phase}\ critical temperature%
\begin{eqnarray}
1/3 &=&\int\limits_{0}^{\infty }\tilde{k}^{2}d\tilde{k}[{\exp}\{(\tilde{k%
}^{2}-\tilde{n}^{-2/3}-\nu /2)/\tilde{T}_{c}\}+1]^{^{-1}}+\int\limits_{0}^{%
\infty }\tilde{K}^{2}d\tilde{K}[{\exp}(\lambda \tilde{K}/2\tilde{T}%
_{c})-1]^{-1}  \nonumber \\
&&-\int\limits_{0}^{\infty }\tilde{K}^{2}d\tilde{K}[{\exp}\{(\lambda 
\tilde{K}+4\nu )/2\tilde{T}_{c}\}-1]^{^{-1}}  \label{working}
\end{eqnarray}
where $\tilde{k}\equiv k/k_{F}$, $\tilde{K}\equiv K/k_{F}$, $\tilde{n}\equiv
n/n_{f}$, $\tilde{T_{c}}\equiv T_{c}/T_{F}$, $\nu \equiv \hbar \omega
_{D}/E_{F},$ and we took $\varepsilon _{K}\simeq \lambda \hbar v_{F}K/4$
[Ref. \cite{Honolulu}, Eq. (12)] for $d=3$. The integrals are exact, the
first and last being expressible as polylogarithm functions $Li_{\sigma }(z)$
or PolyLog[$\sigma $, z]  \cite{39} where
\begin{equation}
-aLi_{\sigma }(-{\normalsize a}z)\equiv {\frac{1}{\Gamma (\sigma )}}
\int_{0}^{\infty }dx{\frac{x^{\sigma -1}}{z^{-1}e^{x}+\mathit{a}}}=-\frac{{1}
}{a}\sum_{l=1}^{\infty }{\frac{(-\mathit{a}z)^{\mathit{l}}}{\mathit{l}
^{\sigma }}}  \label{poly}
\end{equation}%
with $z$ an effective fugacity. For $a=-1$ (\ref{poly}) reduces to the Bose
integral $g_{\sigma }(z)$ which for $z=1$ and $\sigma \geq 1$ becomes the
Riemann Zeta function $\zeta (\sigma )$\ of order $\sigma $; for $a=1$ (\ref%
{poly}) becomes the Fermi integral $f_{\sigma }(z)$. Both integrals are as
defined in Appendices D and E of Ref. \cite{Path}. Since $Li_{\sigma
}(1)\equiv \zeta (\sigma )$, the second integral in (\ref{working}) gives $\zeta (3)$, and the working number
equation simplifies to
\begin{eqnarray}
1/3&=&-\frac{\sqrt{\pi }\tilde{T}_{c}^{\ 3/2}}{4}PolyLog[3/2,{ }-{\exp%
}\{(\nu /2+\tilde{n}^{-2/3})/\tilde{T}_{c}\}] \nonumber \\
&+&\frac{16\tilde{T}_{c}^{\ 3}}{\lambda ^{3}}{ \{}{\zeta (3)-}{{\textit{PolyLog}}}{[3,}{ \exp}
(-2\nu /\tilde{T}_{c}){]}{ \}.}
\end{eqnarray}
This can now be solved for $\tilde{T}_{c}$ as a function of $\tilde{n}$
which is plotted as the dashed curve in Fig. 2 for $\lambda =
{\frac12}$ and $\nu =0.005.$
\vskip -0.0truecm
\begin{figure}
\includegraphics[height=0.55\textheight]{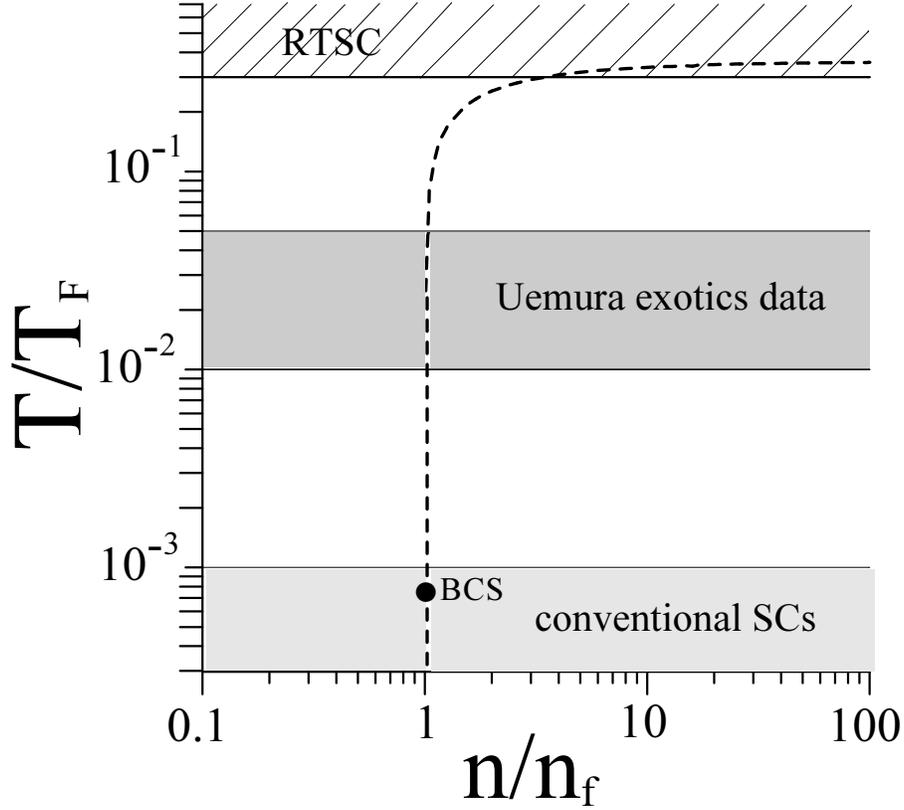}
\caption{Temperature $T$ (in units of $T_{F}$) \textit{vs.} electron density $n$
(in units of $n_{f}$ as defined in text) phase diagram in 3D showing the
critical temperature $T_{c}$ phase boundary for the pure 2e-CP BEC phase
(dashed curve) for $\lambda =$ $1/2$ with $\hbar \omega _{D}/E_{F}=0.005$.
The BCS $T_{c}/T_{F}\simeq 1.134(\hbar \omega _{D}/E_{F})\exp (-1/\lambda )$
gives $\simeq 0.0008$ for these values of $\lambda $ and $\hbar \omega
_{D}/E_{F}$; it is marked by a large dot. There is no pure 2h-CP BEC phase\
solution for $n/n_{f}>1$, nor a mixed phase with both types of BECs for any $%
n/n_{f}>0$, as reported in Ref. \cite{PLA2}\ for $f\neq 0$ but for a
quadratic CP dispersion.\ RTSC refers to room-temperature superconductivity
in a material with $T_{F}=10^{3}$ $K.$}
\end{figure}
\vskip -0.0truecm
The \textit{pure 2h-CP BEC phase }comes from the limiting case $2\mu
=E_{-}(0)$ as then the second, but not the first, relation in (\ref{Emas})
is satisfied, so that, again from (\ref{3conditions}), $N_{0}(T)=0$ for all $%
T$ but $M_{0}(T)\neq 0$ for some $T$. A working number equation similar to but different from (\ref%
{working}) follows and the critical temperature for this phase is now
determined by $m_{0}(T_{c},n)=0.$ It eventually gives
\vskip -0.5truecm
\begin{eqnarray}
1/3&=&-\frac{\sqrt{\pi }\tilde{T}_{c}^{\ 3/2}}{4}PolyLog[3/2,{ }-{\exp%
}\{(\tilde{n}^{-2/3}-\nu /2)/\tilde{T}_{c}\}] \nonumber \\
&+&\frac{16\tilde{T}_{c}^{\ 3}}{\lambda ^{3}}{ \{}\mathit{PolyLog}{[3,}{ \exp}(-2\nu /\tilde{T}_{c})%
{]-\zeta (3)}{ \}.}
\end{eqnarray}
\textit{No bounded solution} of this equation for $\tilde{T}_{c}$ when $\lambda =1/2$
and $\nu \equiv \hbar \omega _{D}/E_{F}=0.005$ was found.
The BCS value from their formula $T_{c}/T_{F}\simeq 1.134(\hbar \omega
_{D}/E_{F})\exp (-1/\lambda )$ is $\simeq $ $0.0008$ (large black dot in
Fig. 2); it lies within the range $T_{c}/T_{F}\approx 10^{-3}$
empirically found for conventional, elemental SCs. Empirical data \cite%
{Uemura} for ``exotic'' SCs, however, fall within the range $%
T_{c}/T_{F}\simeq 0.01-0.05$. Thus, moderate departures from perfect
2e/2h-CP symmetry enable the ICBFM to access, unlike BCS theory, empirical $
T_{c}$ values for exotic SCs \textit{without abandoning electron-phonon
dynamics}.

Finally, for intermediate values of $\mu $, namely for $E_{-}(0)<2\mu
<E_{+}(0),$ neither the first nor second equations of (\ref{3conditions})
are satisfied so that $n_{0}(T)=0=m_{0}(T)$ for all $T$; this implies no
condensed phases whatsoever. Thus, the ICBFM (characterized by zero BF
coupling $f$) \textit{contains no mixed phase} in contrast to the CBFM where 
$f\neq 0$ (Ref. \cite{PLA2} for a quadratic dispersion). This case
will be treated elsewhere with the correct linear dispersion.

\section{CONCLUSIONS}

Cooper pairs (CPs)\ are meaningless if referred to the ideal Fermi gas
``sea'' when hole pairs are included along with electron pairs, but survive
as positive-energy, finite-lifetime, plasmon-like objects with a linear
(instead of quadratic) rise in total, or center-of-mass, momentum $K$ when
referred to a BCS-correlated sea instead. 

The new generalized BEC (GBEC) theory includes as limiting cases the
following theories:\ i) BCS and\ ii) BCS-Bose ``crossover,'' when the BE\
condensate consists of equal numbers of electron- and hole-pairs. It also
contains: iii) the Friedberg-T.D. Lee BEC model,\ iv) the ``ideal
boson-fermion model'' (IBFM), and v) ordinary BEC theory when there are no
unpaired fermions. The BCS condensate is precisely a BE condensate of equal
numbers of 2e/2h-pairs and weak coupling. Without abandoning electron-phonon
dynamics the GBEC theory leads to 2-to-3 order-of-magnitude higher $T_{c}$%
's---including room-temperature superconductivity. All this rests on four
essential ingredients: 1) 2h-CPs cannot and must not\ be neglected in a
fully self-consistent treatment in any many-\textit{fermion }system,
otherwise a spurious value of $T_{c}$ may result that corresponds not to a
stable but rather to a \textit{metastable} state;\ 2) CPs are \textit{bosons}%
, even though BCS pairs not; 3)\ CPs are \textit{linearly-dispersive }for
small $K$; 4) to achieve higher $T_{c}$'s one must depart from the perfect
2e-/2h-CP symmetry of the BCS condensate which in fact is a BE condensate.
Neglected in the GBEC theory thus far, however,\ are: a) $K>0$\ terms in the
boson-fermion vertex interactions; b) boson-boson interactions (as also in
BCS theory); c) a $T>0$\ Bethe-Salpeter CP treatment; d) different hole and
electron effective masses;\ and e) ionic-lattice crystallinity effects which
might initially be introduced via Van Hove singularities in the electronic
DOS or via ``bipolarons'' instead of CPs.

Finally, at least two mysteries have surfaced here. In spite of neglecting
boson-boson interactions between\ severely-overlapping CPs, why has the BCS
theory been so successful in describing at least conventional SCs?\ Why are
simple models (such as the BCS or the GBEC theories) quite able to be of any
relevance whatsoever in such complex strongly-interacting many-electron
systems like SCs?


\begin{theacknowledgments}
We acknowledge UNAM-DGAPA-PAPIIT (Mexico) grant \#
IN108205, and CONACyT (Mexico) grant \# 41302F, for partial support. MdeLl
appreciates the hospitality shown him at the Texas Center for
Superconductivity, University of Houston during a sabbatical leave, and
thanks Paul C.W. Chu, D.M. Eagles, Randy Hulet, Colin Gough and Chin-Sen Ting for
discussions and/or encouragement.
\end{theacknowledgments}



\bibliographystyle{aipproc}   

\bibliography{sample}

\begin{thebibliography}{9}
\bibitem{Blatt} J.M. Blatt, \textit{Theory of Superconductivity,} Academic,
New York, 1964.

\bibitem{BF} M.R. Schafroth, \textit{Phys. Rev.} \textbf{96}, 1442 (1954).

\bibitem{BF1} M.R. Schafroth, S.T. Butler, and J.M. Blatt, \textit{Helv. Phys. Acta} \textbf{30}, 93\ (1957).

\bibitem{BF2} M.R. Schafroth, \textit{Sol. State Phys. }\textbf{10}, 293 (1960).

\bibitem{bcs} J. Bardeen, L.N. Cooper and J.R. Schrieffer, \textit{Phys. Rev.} \textbf{108}, 1175 (1957).

\bibitem{Bog58} N.N. Bogoliubov, \textit{JETP} \textbf{34}, 41 (1958).

\bibitem{bts} N.N. Bogoliubov, V.V. Tolmachev and D.V. Shirkov, \textit{Fortschr. Phys. }\textbf{6}, 605 (1958); and also in \textit{A New Method in
the Theory of Superconductivity,} Consultants Bureau, New York, 1959.

\bibitem{BF3} J. Ranninger, R. Micnas and S. Robaszkiewicz, \textit{Ann. Phys. Fr.} \textbf{%
13}, 455 (1988).

\bibitem{BF5} R. Friedberg and T.D. Lee, \textit{Phys. Rev. B\ }\textbf{40},
6745 (1989).

\bibitem{BF6} R. Friedberg, T.D. Lee, and H.-C. Ren, \textit{Phys. Lett. A} 
\textbf{152}, 417 and 423\textbf{\ }(1991).

\bibitem{BF7a} V.V. Tolmachev, \textit{Phys. Lett. A }\textbf{266}, 400
(2000).

\bibitem{PLA2} M. de Llano and V.V. Tolmachev, \textit{Physica A }\textbf{317%
}, 546 (2003).

\bibitem{CMT02} J. Batle, M. Casas, M. Fortes, M. de Llano, F.J. Sevilla,
M.A. Sol\'{\i}s, and V.V. Tolmachev, \textit{Cond. Matter Theories }\textbf{%
18}, 111 (2003). Cond-mat/0211456.

\bibitem{Frontiers} M. de Llano, ``High-T$_{c}$ Superconductivity via BCS
and BEC Unification: A Review''\ in \textit{Frontiers in Superconductivity
Research}, edited by B.P. Martins, Nova Science Publishers, New York, 2004,
p. 1. Also in cond-mat/0405071.

\bibitem{classical} B.S. Deaver, Jr. and W.M. Fairbank, \textit{Phys. Rev.
Lett.} \textbf{7}, 43 (1961).

\bibitem{classical2} R. Doll and M. N\"{a}bauer, \textit{Phys. Rev. Lett.} 
\textbf{7}, 51 (1961).

\bibitem{cuprates} C.E. Gough, M.S Colclough, E.M. Forgan, R.G. Jordan, M.
Keene, C.M. Muirhead, I.M. Rae, N. Thomas, J.S. Abell, and S. Sutton, 
\textit{Nature}\textbf{\ 326, }855 (1987).

\bibitem{Gough} C.E. Gough, priv. comm.

\bibitem{Coo} L.N. Cooper, \textit{Phys. Rev. }\textbf{104}, 1189 (1956).

\bibitem{PRB2000} S.K. Adhikari, M. Casas, A. Puente, A. Rigo, M. Fortes,
M.A. Sol\'{\i}s, M. de Llano, A.A. Valladares, and O. Rojo, \textit{Phys.
Rev. B} \textbf{62}, 8671 (2000).

\bibitem{PhysicaC} S.K. Adhikari, M. Casas, A. Puente, A. Rigo, M. Fortes,
M. de Llano, M.A. Sol\'{\i}s, A. A. Valladares, and O. Rojo, \textit{Physica
C }\textbf{351}, 341 (2001).

\bibitem{FW} A.L. Fetter and J.D. Walecka, \textit{Quantum Theory of
Many-Particle Systems}, McGraw-Hill, New York, 1971.

\bibitem{PC05} M. de Llano, F.J. Sevilla, M.A. Sol\'{\i}s, and J.J. Valencia%
\textit{, Cond. Mat. Phys.} \textbf{20} (in press) (2005). Also in
cond-mat/0509118.

\bibitem{Bog} N.N. Bogoliubov, \textit{N. Cim.}\textbf{\ 7}, 794 (1958).

\bibitem{Schrieffer} J.R. Schrieffer, \textit{Theory of Superconductivity},
Benjamin, New York, 1964.

\bibitem{ABH} {P.W. Anderson, \textit{Phys. Rev.} }\textbf{112}, 1900{\
(1958).}

\bibitem{Higgs} {P.W. Higgs, \textit{Phys. Lett.} }\textbf{12}, 132 {(1964).}

\bibitem{FortesdeLlano05} M. Fortes, M. de Llano, and M.A. Sol\'{\i}s, to be
published.

\bibitem{ANFdeLl} V.C. Aguilera-Navarro, M. Fortes, and M. de Llano,\textit{%
\ Sol. St. Comm.} \textbf{129}, 577\textbf{\ }(2004).

\bibitem{Honolulu} M. Fortes, M.A. Sol\'{\i}s, M. de Llano, and V.V.
Tolmachev, \textit{Physica C} \textbf{364}, 95 (2001).

\bibitem{Brandow} B.D. Brandow, \textit{Phys. Repts.} \textbf{296}, 1 (1998).

\bibitem{BKT} V.L. Berezinskii, Sov. Phys. \textit{JETP} \textbf{34, }610
(1972).

\bibitem{KT} J.M. Kosterlitz and D.J. Thouless, \textit{J. Phys.} \textbf{C6}%
, 1181 (1973).

\bibitem{organometallics} D. J\'{e}rome, \textit{Science} \textbf{252}, 1509
(1991).

\bibitem{jerome1} J.M. Williams, A.J. Schultz, U. Geiser, K.D. Carlson, A.M.
Kini, H.H. Wang, W.K. Kwok, M.H. Whangbo, and J.E. Schirber, \textit{Science}
\textbf{252}, 1501 (1991).

\bibitem{jerome2} H. Hori, \textit{Int. J. Mod Phys. B} \textbf{8,} 1 (1994).

\bibitem{Peter} G.M. Zhao and P. Beeli, ``Magnetic evidence for hot
superconductivity in multi-walled carbon nanotubes.'' Cond-mat/0509037.

\bibitem{AGD} A.A. Abrikosov, L.P. Gorkov, and I.E. Dzyaloshinskii, \textit{%
Methods of Quantum Field in Statistical Physics},Dover, New York,
1975, \S 33.

\bibitem{Bog47} N.N. Bogoliubov, \textit{J. Phys. (USSR)} \textbf{11}, 23
(1947).

\bibitem{Friedel} J. Labb\'{e}, S. Barisic, and J. Friedel, \textit{Phys.
Rev. Lett.} \textbf{19}, 1039\ (1967).

\bibitem{Eagles} D.M. Eagles, \textit{Phys. Rev. }\textbf{186}, 456 (1969).

\bibitem{BCS-Bose} M. Randeria, in \textit{Bose-Einstein Condensation},
edited by A. Griffin \textit{et al.}, Cambridge University, Cambridge,
UK,1995, p. 355.

\bibitem{LevinPhysReps} Q. Chen, J. Stajic, S. Tan, K. Levin, \textit{Phys.
Repts. }\textbf{412}, 1 (2005).

\bibitem{BF9} M. Casas, N.J. Davidson, M. de Llano, T.A. Mamedov, A. Puente,
R.M. Quick, A. Rigo, and M.A. Sol\'{\i}s, \textit{Physica A}\textbf{\ 295},
146\textbf{\ }(2001).

\bibitem{BF10} M. Casas, M. de Llano, A. Puente, A. Rigo, and M.A. Sol\'{\i}%
s, \textit{Sol. State Comm. }\textbf{123}, 101 (2002).


\bibitem{Migdal} A.B. Migdal, \textit{JETP} \textbf{7}, 996 (1958).

\bibitem{39} S. Wolfram, \textit{The MATHEMATICA Book}, 3rd. Ed. (Wolfram
Media, IL, 1996) p. 743.

\bibitem{Path} R.K. Pathria, \textit{Statistical Mechanics}, 2nd Ed.
(Pergamon, Oxford, 1996).

\bibitem{Uemura} Y.J. Uemura, \textit{J. Phys.: Cond. Matter} \textbf{16},
S4515 (2004). Also in cond-mat/0406301.
\end{thebibliography}

\IfFileExists{\jobname.bbl}{}
 {\typeout{}
  \typeout{******************************************}
  \typeout{** Please run "bibtex \jobname" to optain}
  \typeout{** the bibliography and then re-run LaTeX}
  \typeout{** twice to fix the references!}
  \typeout{******************************************}
  \typeout{}


\end{document}